# Multiple Access Channels with Cooperative Encoders and Channel State Information

Reza Khosravi-Farsani, and Farokh Marvasti

*Advanced Communications Research Institute (ACRI)*
*Department of Electrical Engineering*
*Sharif University of Technology, Tehran, Iran*
*Email: reza.khosravi@ee.sharif.ir, marvasti@sharif.ir*

*Abstract*—**The two-user Multiple Access Channel (MAC) with cooperative encoders and Channel State Information (CSI) is considered where two different scenarios are investigated: A two-user MAC with common message (MACCM) and a two-user MAC with conferencing encoders (MACCE). For both situations, the two cases where the CSI is known to the encoders either non-causally or causally are studied. Achievable rate regions are established for both discrete memoryless channels and Gaussian channels with additive interference. The achievable rate regions derived for the Gaussian models with additive interference known non-causally to the encoders are shown to coincide with the capacity region of the same channel with no interference. Therefore, the capacity region for such channels is established.**

## I. INTRODUCTION

THE Multiple Access Channel (MAC) is a communication system in which a number of transmitters send information to a receiver. In a MAC, the transmitters may also cooperate with each other by exchanging information. The MAC with correlated sources was first investigated in [1] wherein the capacity region of the discrete two-user MAC with cooperative encoders via common information was established. In fact, it was shown that the superposition encoding technique achieves the capacity region. The two-user Gaussian MAC with common message (MACCM) was studied in [2] (also see [3] for recent developments). In another scenario, the transmitters exchange information via conferencing. The two-user MAC with conferencing encoders (MACCE) was first introduced by Willems [4]. In this system, two transmitters are connected by communication links of finite capacity, allowing the encoders to communicate over noise-free bit pipes of the given capacity. The capacity region of the discrete two-user MACCE was determined in [4], and the Gaussian model was studied in [5].

Channels with Channel State Information (CSI) available at the transmitter were studied initially by Shannon [6], where he characterized the capacity of a single-user channel with CSI causally known to the transmitter. The capacity of the single-user channel with CSI in which the transmitter has access to the CSI non-causally was determined in [7]. In 1983 [8], Costa considered the Gaussian counterpart of the single-user channel with non-causal CSI, i.e. a Gaussian channel with additive interference noise which is known non-causally to the transmitter. Costa showed that for this system coined "writing on dirty paper", the capacity is the same as when there is no interference, i.e., the additive interference known non-causally to the transmitter has no penalty on the capacity. The multiuser channels with causal and non-causal CSI were also studied in several papers [9-12]; specially, the two-user MAC with causal and non-causal CSI was studied in [9, 12]. Recently, the two-user MAC with CSI and with partially cooperating encoders where cooperation is utilized to generate empirical state coordination between the encoders as well as to share information about the private messages of the users, has been studied in [13]. Also, the Gaussian MACCE with two independent additive interferences has been studied [14] wherein the author has established the capacity region to within a constant gap. Costa's result for dirty paper channel was also generalized to some multiuser systems; specifically, the same result was established for the two-user MAC [15, 16], for the broadcast channel and the relay channel [16], for the relay broadcast channel [17] and for the two-user MACCE [5].

In this paper, we investigate the two-user MAC with cooperative encoders and with CSI. Two different situations are considered: A two-user MACCM and a two-user MACCE. For both situations, we study the two cases where the CSI is known to the encoders either non-causally or causally. The decoder is assumed to have no access to the CSI. It should be noted that unlike [13], in our model for the MACCE with CSI both encoders have access to the same CSI (perfect CSI). Hence, cooperation is only utilized to share information about the private messages of the users and indeed for this purpose a conferencing is held between the two encoders before the transmission begins. We establish achievable rate regions for the discrete memoryless channels and the Gaussian channels with additive interference modeled as the channel state.

Furthermore, we show that the achievable rate regions derived for the Gaussian models with additive interference known non-causally to both transmitters coincide with the capacity region of the underlying channel with no interference, therefore yielding the capacity region for these important cases. Our capacity results for the Gaussian MAC with cooperative encoders and with additive interference known non-causally to both encoders, generalize the previous known results for the two-user multiple access Costa's type channel, i.e. the two-user MAC without common message [14, 15], the two-user MACCE [5], and the-two-user MAC with degraded message sets [18].

The rest of the paper is organized as follows: Channel models and definitions are given in Section II. The main results are stated in Section III, where the two-user MACCM with CSI is studied in Section III-A, and the two-user MACCE with CSI is studied in Section III-B. Finally, the paper is concluded in Section IV.

## II. CHANNEL MODELS AND DEFINITIONS

In this paper, the following notations are used: Random variables (r.v.) are denoted by capital letters (e.g. $X$) and lower case letters are used to show their realization (e.g. $x$). The probability distribution function (p.d.f.) of a r.v. $X$ with alphabet set $\mathcal{X}$ is denoted by $P_X(x)$ where $x \in \mathcal{X}$ and $P_{X|Y}(x|y)$ stands for the conditional p.d.f. of $X$ given $Y$. $\mathbb{E}[.]$, and $\mathbb{h}(.)$ denote the expectation operator and differential entropy, respectively. Finally, the set of real numbers is denoted by $\mathbb{R}$.

*Definition:* A discrete two-user MAC with CSI denoted by $\langle \mathcal{X}_1, \mathcal{X}_2, \mathcal{Y}, \mathcal{S}, P_S(s), \mathbb{P}(y|x_1, x_2, s)\rangle$, is a channel with two input alphabets $\mathcal{X}_1, \mathcal{X}_2$, an output alphabet $\mathcal{Y}$. The input and output alphabets are finite sets. The state of the channel $S$ ranges over the state space $\mathcal{S}$ (finite set) according to the known p.d.f. $P_S(s)$. The channel is described by the transition p.d.f. $\mathbb{P}(y|x_1, x_2, s)$. The channel and also the state process are assumed to be memoryless, i.e., for $n \geq 1$:

$$P(y^n|x_1^n, x_2^n, s^n) = \prod_{t=1}^{n} \mathbb{P}\big(y_t|x_{1,t}, x_{2,t}, s_t\big) \ ,$$
$$P(s^n) = \prod_{t=1}^{n} P_S(s_t) \tag{1}$$

In this paper, we study the two-user MAC with cooperative encoders and with CSI where two different cases are considered:

*A) The two-user MACCM with CSI:* In this network scenario each encoder sends a private message over the channel and both encoders cooperate to transmit a common message. Fig. 1 illustrates the channel.

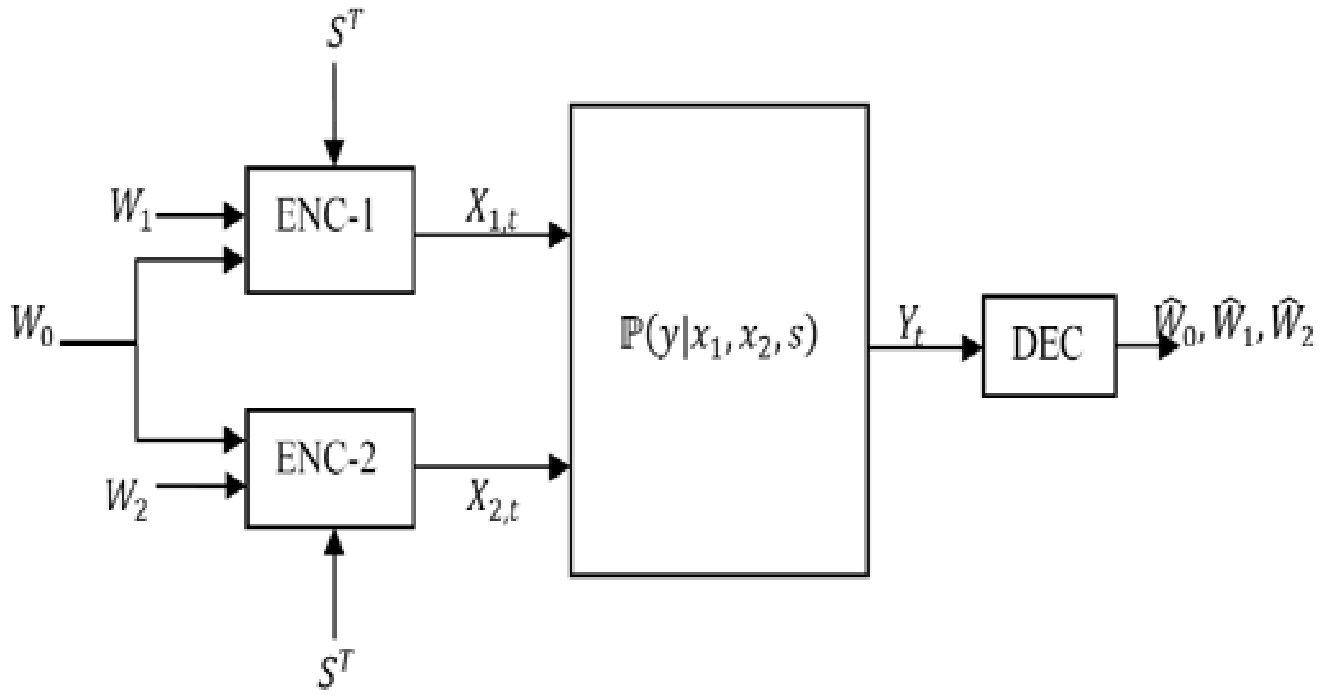

Figure 1. The two-user MACCM with CSI: For $T = n$ the two encoders have access to the CSI non-causally, while for $T = t$ the two encoders have access to the CSI causally.

*Encoding and decoding:* For the two-user MACCM with CSI in Fig.1, a length-$n$ code $\mathbb{C}^n(R_0, R_1, R_2)$ with two private message sets $\mathcal{W}_i = [1, \dots, 2^{nR_i}], i = 1,2,$ and a common message set $\mathcal{W}_0 = [1, \dots, 2^{nR_0}]$, consists of two sets of encoder functions $\{\mathfrak{E}_{i,t}(.)\}_{t=1}^{n}, i = 1,2$, and a decoder function $\mathfrak{D}(.)$ which are defined as follows:

For the case that the encoders have access to the CSI non-causally, the encoder functions are given by:

$$\begin{cases} \mathfrak{E}_{1,t}(.): \mathcal{W}_1 \times \mathcal{W}_0 \times \mathcal{S}^n \to \mathcal{X}_1 \\ \mathfrak{E}_{2,t}(.): \mathcal{W}_2 \times \mathcal{W}_0 \times \mathcal{S}^n \to \mathcal{X}_2 \end{cases}, \qquad t = 1, \dots, n$$

which generate $X_{i,t} = \mathfrak{E}_{i,t}(W_i, W_0, S^n)$, $i = 1,2$; while for the case that the encoders have access to the CSI causally, the encoder functions are given by:

$$\begin{cases} \mathfrak{E}_{1,t}(.): \mathcal{W}_1 \times \mathcal{W}_0 \times \mathcal{S}^t \to \mathcal{X}_1 \\ \mathfrak{E}_{2,t}(.): \mathcal{W}_2 \times \mathcal{W}_0 \times \mathcal{S}^t \to \mathcal{X}_2 \end{cases}, \qquad t = 1, \dots, n$$

which generate $X_{i,t} = \mathfrak{E}_{i,t}(W_i, W_0, S^t)$, $i = 1,2$; the decoder function $\mathfrak{D}(.)$ for both non-causal and causal CSI is given by:

$$\mathfrak{D}(.): \mathcal{Y}^n \to \mathcal{W}_0 \times \mathcal{W}_1 \times \mathcal{W}_2$$

which estimates the messages $W_0, W_1, W_2$ as: $\big(\widehat{W}_0, \widehat{W}_1, \widehat{W}_2\big) = \mathfrak{D}(Y^n)$. The rate of the code is the triple $(R_0, R_1, R_2)$. The average error probability of decoding $P_e^{\mathbb{C}^n}$ is given by:

$$P_e^{\mathbb{C}^n} := \frac{1}{2^{n(R_0+R_1+R_2)}} \sum_{w_0, w_1, w_2} Pr\left(\begin{matrix}\big(\widehat{W}_0, \widehat{W}_1, \widehat{W}_2\big) \neq \\ (w_0, w_1, w_2)\end{matrix} \middle| (w_0, w_1, w_2) \text{ is sent}\right)$$

A triple of non-negative numbers $(R_0, R_1, R_2)$ is said to be achievable for the two-user MACCM with CSI if for every $\epsilon > 0$ and for all $n$ sufficiently large, there exists a length-$n$ code $\mathbb{C}^n(R_0, R_1, R_2)$ such that $P_e^{\mathbb{C}^n} \leq \epsilon$. The capacity region of the channel is the closure of all achievable rates.

We also investigate the two-user Gaussian MACCM in the presence of additive interference $S$ known non-causally or causally to the transmitters. The channel is defined as follows:

$$Y_t = X_{1,t} + X_{2,t} + S_t + Z_t, \qquad t = 1, \dots, n \tag{2}$$

where $\{X_{i,t}\}_{t=1}^{n}, i=1,2,$ and $\{Y_t\}_{t=1}^{n}$ are the $\mathbb{R}$-valued transmitted and received signals, respectively. The process $\{S_t\}_{t=1}^{n}$ is the additive interference known non-causally or causally to both transmitters, which is assumed to be Independent Identically Distributed (i.i.d.) such that $S_t, t=1,\dots,n$ has Gaussian p.d.f. with zero mean and variance $P_S$. $\{Z_t\}_{t=1}^{n}$ is Additive White Gaussian Noise (AWGN) with zero mean and variance $P_Z$, which is independent of $\{S_t\}_{t=1}^{n}$.

Encoding and decoding for the Gaussian channel in (2) is defined similar to the discrete channel model. Also, an average power constraint is imposed on the codewords of each encoder, and the encoder $i$ is subject to an average power constraint $P_i \in \mathbb{R}^+$, i.e.,

$$\frac{1}{n}\mathbb{E}\left[\sum_{t=1}^{n} X_{i,t}^2\right] \leq P_i\,, \qquad i=1,2 \tag{3}$$

*B) The two-user MACCE with CSI:* In this network scenario, two private messages are transmitted over the channel by two cooperative encoders which have been connected to each other by links of finite capacities $C_{12}$ and $C_{21}$. Each encoder cooperates with the other encoder by holding a conference before the transmission begins. Fig. 2 illustrates the channel.

*Encoding and decoding:* For the two-user MACCE with CSI in Fig. 2, a length-$n$ code $\mathbb{C}^n(R_1,R_2,C_{12},C_{21})$ with two private message sets $\mathcal{W}_i=\{1,\dots,2^{nR_i}\}, i=1,2$, is defined as follows:

*Conferencing before the transmission:* The definition of conferencing between the encoders is the same as [4]. The code $\mathbb{C}^n(R_1,R_2,C_{12},C_{21})$ consists of two sets of $L$ conferencing functions $\{\psi_{i,1}(.),\dots,\psi_{i,L}(.)\}$, $i=1,2$, and two sets of conferencing (finite) alphabets $\{\mathcal{V}_{i,1},\dots,\mathcal{V}_{i,L}\}$, $i=1,2$. The conferencing functions $\psi_{i,l}(.)$, $i=1,2$, $l=1,\dots,L$, are given by:

$$\psi_{1,l}(.): \mathcal{W}_1 \times \mathcal{V}_2^{l-1} \to \mathcal{V}_{1,l}\,, \qquad V_{1,l}=\psi_{1,l}(W_1,V_2^{l-1})$$

$$\psi_{2,l}(.): \mathcal{W}_2 \times \mathcal{V}_1^{l-1} \to \mathcal{V}_{2,l}\,, \qquad V_{2,l}=\psi_{2,l}(W_2,V_1^{l-1})$$

A conference is said to be $(C_{12},C_{21})$-admissible [4] if the sets of conferencing functions are such that:

$$\sum_{l=1}^{L}\log\|\mathcal{V}_{1,l}\| \leq nC_{12}\,, \sum_{l=1}^{L}\log\|\mathcal{V}_{2,l}\| \leq nC_{21}$$

where $\|A\|$ denotes the cardinality of the set $A$.

After the conferencing, the encoders 1 and 2 know the sequences $V_2^L=(V_{2,1},\dots,V_{2,L})$ and $V_1^L=(V_{1,1},\dots,V_{1,L})$, respectively. The code $\mathbb{C}^n(C_{12},C_{21})$ consists of two sets of encoder functions $\{\mathfrak{E}_{i,t}(.)\}_{t=1}^{n}, i=1,2$, and a decoder function $\mathfrak{D}(.)$ defined as follows:

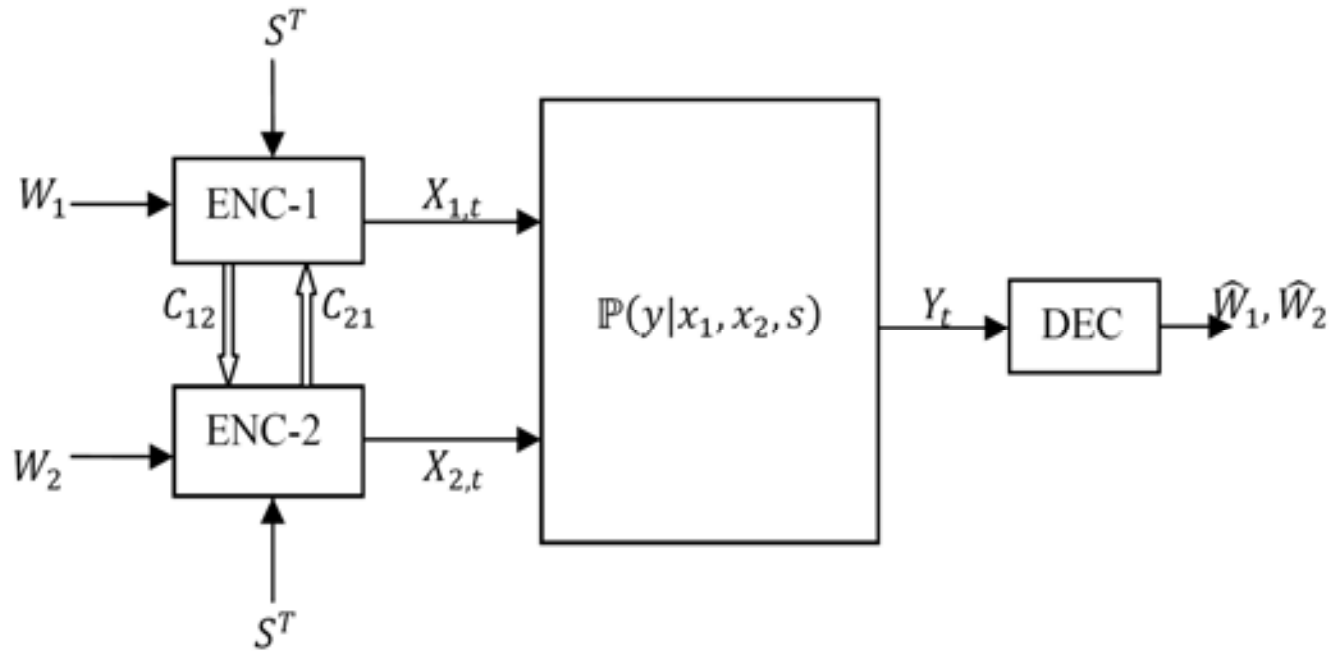

Figure 2. The two-user MACCE with CSI: For $T=n$ the two encoders have access to the CSI non-causally, while for $T=t$ the two encoders have access to the CSI causally.

For the case that the encoders have access to the CSI non-causally, the encoder functions are given by:

$$\begin{cases}\mathfrak{E}_{1,t}(.): \mathcal{W}_1 \times \mathcal{V}_2^L \times \mathcal{S}^n \to \mathcal{X}_1 \\ \mathfrak{E}_{2,t}(.): \mathcal{W}_2 \times \mathcal{V}_1^L \times \mathcal{S}^n \to \mathcal{X}_2\end{cases}, \qquad t=1,\dots,$$

which generate $X_{1,t}=\mathfrak{E}_{1,t}(W_1,V_2^L,S^n)$ and $X_{2,t}=\mathfrak{E}_{2,t}(W_2,V_1^L,S^n)$; while for the case that the encoders have access to the CSI causally, the encoder functions are given by:

$$\begin{cases}\mathfrak{E}_{1,t}(.): \mathcal{W}_1 \times \mathcal{V}_2^L \times \mathcal{S}^t \to \mathcal{X}_1 \\ \mathfrak{E}_{2,t}(.): \mathcal{W}_2 \times \mathcal{V}_1^L \times \mathcal{S}^t \to \mathcal{X}_2\end{cases}, \qquad t=1,\dots,n$$

which generate $X_{1,t}=\mathfrak{E}_{1,t}(W_1,V_2^L,S^t)$ and $X_{2,t}=\mathfrak{E}_{2,t}(W_2,V_1^L,S^t)$; the decoder function $\mathfrak{D}(.)$ for both causal and noncausal CSI is defined as the MACCM. The rate of the code is the pair $(R_1,R_2)$.

The definitions of the error probability of the code and also the capacity region for the two-user MACCE with CSI are similar to the MACCM and are omitted. The two-user Gaussian MACCE with additive interference is also defined by (2) with the power constraints (3), where the encoding/decoding and the conferencing schemes for this channel are defined in the same way as the discrete model which are omitted.

In the next section, we state our main results for the channel models defined here.

## III. Main Results

***III-A) The two-user MACCM with CSI:***

In this section, we state and prove our main results for the two-user MACCM with CSI. We first consider the discrete memoryless channel, and then we will consider the Gaussian model.

In the following, we establish achievable rate regions for the discrete channels with non-causal and causal CSI:

***Theorem 1 (Discrete MACCM):*** Consider the two-user discrete memoryless MACCM with CSI:

**I)** For the case where the CSI is known non-causally to both transmitters, the following rate region is achievable:

$$\bigcup_{\substack{P_{U|S}P_{V_1|US}P_{V_2|US}\\ X_i=f_i(V_i,U,S),i=1,2}} \left\{ \begin{array}{l} 0 \le R_0, R_1, R_2 \\ R_1 \le I(V_1;Y|V_2,U) - I(V_1;S|V_2,U) \\ R_2 \le I(V_2;Y|V_1,U) - I(V_2;S|V_1,U) \\ R_1 + R_2 \le I(V_1,V_2;Y|U) - I(V_1,V_2;S|U) \\ R_0 + R_1 + R_2 \le I(U,V_1,V_2;Y) \\ \qquad\qquad\qquad -I(U,V_1,V_2;S) \end{array} \right\} \tag{4}$$

where $U, V_1, V_2$ are three auxiliary r.v.'s with finite ranges $\mathcal{U}, \mathcal{V}_1, \mathcal{V}_2$, respectively, and the joint p.d.f. of the r.v.'s $S, U, X_1, V_1, X_2, V_2, Y$ is given by:

$$P_{SUX_1V_1X_2V_2Y}(s,u,x_1,v_1,x_2,v_2,y) =$$
$$P_S(s)P_{U|S}(u|s)P_{X_1V_1|US}(x_1,v_1|u,s)P_{X_2V_2|US}(x_2,v_2|u,s) \times$$
$$\mathbb{P}(y|x_1,x_2,s) \tag{5}$$

where the distribution $P_{X_i|V_iUS}(x_i|v_i,u,s), i = 1,2$, is such that it takes its values from the set $\{0,1\}$; equivalently, $X_i = f_i(V_i,U,S)$, where $f_i(.), i = 1,2$, is an arbitrary deterministic function.

**II)** For the case where the CSI is known causally to both transmitters, the following rate region is achievable:

$$\bigcup_{\substack{P_U P_{V_1|U} P_{V_2|U}\\ X_i=f_i(V_i,U,S),i=1,2}} \left\{ \begin{array}{l} 0 \le R_0, R_1, R_2 \\ R_1 \le I(V_1;Y|V_2,U) \\ R_2 \le I(V_2;Y|V_1,U) \\ R_1 + R_2 \le I(V_1,V_2;Y|U) \\ R_0 + R_1 + R_2 \le I(U,V_1,V_2;Y) \end{array} \right\} \tag{6}$$

where $U, V_1, V_2$ are three auxiliary r.v.'s with finite ranges $\mathcal{U}, \mathcal{V}_1, \mathcal{V}_2$, respectively, and the joint p.d.f. of the r.v.'s $S, U, X_1, V_1, X_2, V_2, Y$ is given by:

$$P_{SUX_1V_1X_2V_2Y}(s,u,x_1,v_1,x_2,v_2,y) =$$
$$P_S(s)P_U(u)P_{V_1|U}(v_1|u)P_{V_2|U}(v_2|u) \times$$
$$P_{X_1|V_1US}(x_1|v_1,u,s)P_{X_2|V_2US}(x_2|v_2,u,s)\mathbb{P}(y|x_1,x_2,s) \tag{7}$$

where the distribution $P_{X_i|V_iUS}(x_i|v_i,u,s), i = 1,2$, is such that it takes its values from the set $\{0,1\}$; equivalently, $X_i = f_i(V_i,U,S)$, where $f_i(.), i = 1,2$, is an arbitrary deterministic function.

***Proof of Theorem 1:***

*Part I)* Refer to [20].

*Part II)* The expression of the achievable rate region given by (6) for the channel with causal CSI can be interpreted as a special case of Part I, where $U, V_1, V_2$ are independent of $S$. Indeed, for coding with causal CSI since $U, V_1, V_2$ are independent of $S$, the encoding scheme does not include binning. At each transmitter, the common message and the private message are encoded in a superposition fashion, such that the common message is encoded using a codeword constructed based on $P_U(u)$ and served as the *cloud center*, and the private message is encoded using a codeword constructed based on $P_{V_i|U}(v_i|u), i = 1,2$, and served as the *satellite*. The transmitter $i, i = 1,2$, then sends $X_i = f_i(V_i,U,S)$ over the channel. A jointly typical decoding scheme is also used at the receiver. ■

Now, we deal with the Gaussian MACCM defined by (2). It is worth noting that the Gaussian channels with additive interference known non-causally to the transmitters, is usually called the "dirty paper channels" [8]. In the next theorem, we prove that the capacity region of the dirty paper MACCM is the same as the clean MACCM wherein the interference (known noise) does not exist. In many situations the transmitters have access to the interference sequence causally. These channels sometimes are called "dirty tape channels" [21]. The capacity of the single-user dirty tape channel is still an open problem; however, achievable rates were established for this channel in [21] and [22]. Recently, in [23] achievable rate regions were established for the multiuser dirty tape channels. In the following, we use the same approach as in [23] and obtain an achievable rate region for the two-user dirty tape MACCM.

Note that, we proved the rate regions given in (4) and (6) for the discrete channel model with non-causal and casual CSI, respectively; however, one can exploit these results to obtain achievable rate regions for the Gaussian setting in (2). Thus, we have the following theorem:

***Theorem 2 (Gaussian MACCM):*** Consider the two-user Gaussian MACCM with additive interference (2). Denote $\Lambda(x) := \frac{1}{2}\log(1+x)$, and $\bar{x} := 1 - x$.

**I)** For the case where the additive interference is known non-causally to both transmitters the capacity region is the same as the capacity when there is no interference and is given by:

$$\bigcup_{0\le\beta_1,\beta_2\le 1} \left\{ \begin{array}{l} 0 \le R_0, R_1, R_2 \\ R_1 \le \Lambda\left(\frac{\beta_1 P_1}{P_Z}\right), \qquad R_2 \le \Lambda\left(\frac{\beta_2 P_2}{P_Z}\right) \\ R_1 + R_2 \le \Lambda\left(\frac{\beta_1 P_1 + \beta_2 P_2}{P_Z}\right) \\ R_0 + R_1 + R_2 \le \Lambda\left(\frac{P_1 + P_2 + 2\sqrt{P_1 P_2 \bar{\beta}_1 \bar{\beta}_2}}{P_Z}\right) \end{array} \right\} \tag{8}$$

**II)** For the case where the additive interference is known causally to both transmitters the following rate region is achievable:

$$\bigcup_{\substack{0\le\beta_1,\beta_2\le 1\\ \alpha_i\in\left[-\sqrt{\frac{P_i}{P_S}},\sqrt{\frac{P_i}{P_S}}\right],\\ i=1,2}} \left\{ \begin{array}{l} 0 \le R_0, R_1, R_2 \\ R_1 \le \Lambda\left(\frac{\beta_1 \tilde{P}_1}{P_Z + (1-\alpha_1-\alpha_2)^2 P_S}\right) \\ R_2 \le \Lambda\left(\frac{\beta_2 \tilde{P}_2}{P_Z + (1-\alpha_1-\alpha_2)^2 P_S}\right) \\ R_1 + R_2 \le \Lambda\left(\frac{\beta_1 \tilde{P}_1 + \beta_2 \tilde{P}_2}{P_Z + (1-\alpha_1-\alpha_2)^2 P_S}\right) \\ R_0 + R_1 + R_2 \le \Lambda\left(\frac{\tilde{P}_1 + \tilde{P}_2 + 2\sqrt{\tilde{P}_1 \tilde{P}_2 \bar{\beta}_1 \bar{\beta}_2}}{P_Z + (1-\alpha_1-\alpha_2)^2 P_S}\right) \end{array} \right\}$$

$$(9)$$

where $\tilde{P}_i := P_i - \alpha_i^2 P_S$.

***Remarks:***

1. By setting $R_2 = 0$ and $\beta_2 = 0$ in the rate region given by (8), we obtain the capacity region of the two-user Gaussian MAC with degraded message sets and with additive interference known non-causally to both transmitters [18].
2. By setting $R_0 = 0$ and $\beta_1 = \beta_2 = 1$ in the rate region given by (8), we obtain the capacity region of the two-user Gaussian MAC without common message with additive interference known non-causally to both transmitters [14], [15].

***Proof of Theorem 2:***

*Part I)* To prove Part I, we exploit the achievable rate region (4). Let $\tilde{U}, \tilde{V}_1, \tilde{V}_2$ be three independent Gaussian distributed r.v.'s with zero mean and unit variance. Also, assume that $\tilde{U}, \tilde{V}_1, \tilde{V}_2$ are independent of the Gaussian interference $S$. Let $\beta_i \in [0,1], i = 1,2$, be two arbitrary real numbers. Define the r.v.'s $U, V_1, V_2, X_1, X_2$ as follows:

$$\begin{cases} X_1 := \sqrt{\bar{\beta}_1 P_1}\tilde{U} + \sqrt{\beta_1 P_1}\tilde{V}_1 \\ X_2 := \sqrt{\bar{\beta}_2 P_2}\tilde{U} + \sqrt{\beta_2 P_2}\tilde{V}_2 \\ U := \tilde{U} + \alpha S, \qquad \alpha := \dfrac{\sqrt{\bar{\beta}_1 P_1} + \sqrt{\bar{\beta}_2 P_2}}{P_1 + P_2 + 2\sqrt{P_1 P_2 \bar{\beta}_1 \bar{\beta}_2} + P_Z} \\ V_1 := \tilde{V}_1 + \gamma_1 S, \qquad \gamma_1 := \dfrac{\sqrt{\beta_1 P_1}}{P_1 + P_2 + 2\sqrt{P_1 P_2 \bar{\beta}_1 \bar{\beta}_2} + P_Z} \\ V_2 := \tilde{V}_2 + \gamma_2 S, \qquad \gamma_2 := \dfrac{\sqrt{\beta_2 P_2}}{P_1 + P_2 + 2\sqrt{P_1 P_2 \bar{\beta}_1 \bar{\beta}_2} + P_Z} \end{cases} \tag{10}$$

*Lemma 1:* By definitions (10), the following equalities hold:

$$\begin{cases} I(V_1;Y|V_2,U) = I(V_1;Y,S|V_2,U) \\ \qquad = I(V_1;Y|V_2,U,S) + I(V_1;S|V_2,U) \\ I(V_2;Y|V_1,U) = I(V_2;Y,S|V_1,U) \\ \qquad = I(V_2;Y|V_1,U,S) + I(V_2;S|V_1,U) \\ I(V_1,V_2;Y|U) = I(V_1,V_2;Y,S|U) \\ \qquad = I(V_1,V_2;Y|U,S) + I(V_1,V_2;S|U) \\ I(U,V_1,V_2;Y) = I(U,V_1,V_2;Y,S) \\ \qquad = I(U,V_1,V_2;Y|S) + I(U,V_1,V_2;S) \end{cases} \tag{11}$$

*Proof of Lemma 1:* Refer to [20].

Now by substituting the r.v.'s. $U, V_1, V_2, X_1, X_2$ as defined in (10) in the rate region (4), and using the equalities (11), we obtain the achievability of (8). Furthermore, the rate region of (8) is the capacity region when there is no interference [3, 18]; hence it is also an upper bound on the capacity. In fact, the rate region (8) is the capacity when the CSI, i.e., the additive interference $S$, is also known at the receiver.

*Part II)* To obtain the achievability of (9), we make use of Part II of Theorem 1 and also the approach of [23]. Let $U, V_1, V_2$ be three independent Gaussian distributed r.v.'s with zero mean and unit variance. Also, assume that $U, V_1, V_2$ are independent of the Gaussian interference $S$. Let also $\beta_i \in [0,1]$ and $\alpha_i \in \left[-\sqrt{\frac{P_i}{P_S}}, \sqrt{\frac{P_i}{P_S}}\right], i = 1,2$, be arbitrary real numbers. Define the r.v.'s $X_1, X_2$ as follows:

$$X_i := \sqrt{\bar{\beta}_i \tilde{P}_i}U + \sqrt{\beta_i \tilde{P}_i}V_i - \alpha_i S, \qquad i = 1,2 \tag{12}$$

By substituting the r.v.'s $U, V_1, V_2, X_1, X_2$ in the rate region (6), we obtain the achievability of (9). In fact to achieve the rate region (9) the encoder $i$, $i = 1,2$, expends a part of its power, i.e., $\alpha_i^2 P_S$, for partly cleaning the interference $S$ from the channel and uses the rest of it, i.e., $P_i - \alpha_i^2 P_S$, to transmit its private message and also the common message. This completes the proof. ■

### *III-B) The two-user MACCE with CSI:*

Now consider the two-user MACCE with CSI. We first establish achievable rate regions for the discrete memoryless channels with non-causal and causal CSI and then extend the results to the Gaussian channels with additive interference.

***Theorem 3 (Discrete MACCE):*** Consider the two-user discrete memoryless MACCM with CSI:

**I)** For the case where the CSI is known non-causally to both transmitters, the following rate region is achievable:

$$\bigcup_{\substack{P_{U|S}P_{V_1|US}P_{V_2|US} \\ X_i = f_i(V_i,U,S), i=1,2}} \left\{ \begin{array}{l} 0 \le R_1, R_2 \\ R_1 \le I(V_1;Y|V_2,U) - I(V_1;S|V_2,U) + C_{12} \\ R_2 \le I(V_2;Y|V_1,U) - I(V_2;S|V_1,U) + C_{21} \\ R_1 + R_2 \le I(V_1,V_2;Y|U) - I(V_1,V_2;S|U) \\ \qquad\qquad + C_{12} + C_{21} \\ R_1 + R_2 \le I(U,V_1,V_2;Y) - I(U,V_1,V_2;S) \end{array} \right\} \tag{13}$$

where $U, V_1, V_2$ are three auxiliary r.v.'s with finite ranges $\mathcal{U}, \mathcal{V}_1, \mathcal{V}_2$, respectively, and the joint p.d.f. of the r.v.'s $S, U, X_1, V_1, X_2, V_2, Y$ is given in the same way as Part I of Theorem 1.

**II)** For the case where the CSI is known causally to both transmitters, the following rate region is achievable:

$$\bigcup_{\substack{P_U P_{V_1|U} P_{V_2|U} \\ X_i = f_i(V_i,U,S), i=1,2}} \left\{ \begin{array}{l} 0 \le R_1, R_2 \\ R_1 \le I(V_1;Y|V_2,U) + C_{12} \\ R_2 \le I(V_2;Y|V_1,U) + C_{21} \\ R_1 + R_2 \le I(V_1,V_2;Y|U) + C_{12} + C_{21} \\ R_1 + R_2 \le I(U,V_1,V_2;Y) \end{array} \right\} \tag{14}$$

where $U, V_1, V_2$ are three auxiliary r.v.'s with finite ranges $\mathcal{U}, \mathcal{V}_1, \mathcal{V}_2$, respectively, and the joint p.d.f. of the r.v.'s $S, U, X_1, V_1, X_2, V_2, Y$ is given in the same way as Part II of Theorem 1.

***Proof of Theorem 3:***

*Part I)* To prove the achievability of (13), we utilize the Willems' approach [4] for the two-user MACCE with CSI. Consider a length-$n$ code $\mathbb{C}^{n}(R_1,R_2,C_{12},C_{21})$ with private messages $W_1\in[1,\dots,2^{nR_1}]$ and $W_2\in[1,\dots,2^{nR_2}]$ for the transmitters 1 and 2, respectively. Define $\hat{R}_1:=min\{R_1,C_{12}\}$ and $\hat{R}_2:=min\{R_2,C_{21}\}$. The set $[1,\dots,2^{nR_i}], i=1,2,$ is partitioned (in a deterministic procedure) into $2^{n\hat{R}_i}$ cells each containing $2^{n(R_i-\hat{R}_i)}$ elements. Let $c_i\in\left[1,\dots,2^{n\hat{R}_i}\right], i=1,2,$ denote the cells labels and $a_i\in\left[1,\dots,2^{n(R_i-\hat{R}_i)}\right], i=1,2,$ denote the elements inside each cell. Define $c_i(w_i):=c_i, i=1,2,$ if $w_i$ is inside the cell $c_i$.

Since $\hat{R}_1\le C_{12}$ and $\hat{R}_2\le C_{21}$, by holding a $(C_{12},C_{21})$-admissible conference, the encoder 1 is able to send $c_1(w_1)\in\left[1,\dots,2^{n\hat{R}_1}\right]$ to the encoder 2 and the encoder 2 is able to send $c_2(w_2)\in\left[1,\dots,2^{n\hat{R}_2}\right]$ to the encoder 1, reliably. Therefore, after holding a $(C_{12},C_{21})$-admissible conference, the channel can be viewed as a two-user MAC with the following common message for transmitters:

$$w_0:=\big(c_1(w_1),c_2(w_2)\big)$$

Note that $a_1$ and $a_2$ remain still unknown for the encoders 2 and 1, respectively. Indeed, after conferencing, $w_0\in\left[1,\dots,2^{n(\hat{R}_1+\hat{R}_2)}\right]$ can be considered as the common message and $a_i\in\left[1,\dots,2^{n(R_i-\hat{R}_i)}\right], i=1,2,$ as the private messages. Therefore, by utilizing the achievable rate region (4) derived in Part I of Theorem 1 for the MACCM, we conclude that the rate pair $(R_1,R_2)$ is achievable for the MACCE provided that:

$$\left\{\begin{array}{l}0\le R_1,R_2\\ R_1-\hat{R}_1\le I(V_1;Y|V_2,U)-I(V_1;S|V_2,U)\\ R_2-\hat{R}_2\le I(V_2;Y|V_1,U)-I(V_2;S|V_1,U)\\ R_1-\hat{R}_1+R_2-\hat{R}_2\le I(V_1,V_2;Y|U)-I(V_1,V_2;S|U)\\ \big(\hat{R}_1+\hat{R}_2\big)+R_1-\hat{R}_1+R_2-\hat{R}_2\le I(U,V_1,V_2;Y)\\ \qquad\qquad\qquad\qquad\qquad\qquad -I(U,V_1,V_2;S)\end{array}\right\} \tag{15}$$

where the joint p.d.f.'s of the r.v.'s is given by (5). Now, the achievability of the rate region (13) for the channel with conferencing encoders is readily obtained from (15) by considering $\hat{R}_1:=min\{R_1,C_{12}\}$ and $\hat{R}_2:=min\{R_2,C_{21}\}$.

*Part II)* Similar to Part I the achievability of (14) is obtained using the Willems' approach [4] to transform the channel with conferencing encoders to a channel with common message, where the consequence of Part II of Theorem 1 is also applied. The details will be omitted. ■

Now, consider the two-user Gaussian MACCE (2). As mentioned before, in [5] the capacity region of this channel for the case in which the interference is known non-causally to both transmitters was established, where it was shown that the known interference could be canceled without any rate penalty. In the following, we present an alternative proof for this result where we directly apply the result of Part I of Theorem 3 and utilize a similar approach to the derivation of Part I of Theorem 2. We also establish an achievable rate region for the situation in which the transmitters have access to the interference causally.

***Theorem 4 (Gaussian MACCE):*** Consider the two-user Gaussian MACCE with additive interference (2). Denote $\Lambda(x):=\frac{1}{2}\log(1+x)$, and $\bar{x}:=1-x$.

**I)** For the case where the additive interference is known non-causally to both transmitters, the capacity region is the same as the capacity when there is no interference and is given by:

$$\bigcup_{0\le\beta_1,\beta_2\le1}\left\{\begin{array}{l}0\le R_1,R_2\\ R_1\le\Lambda\left(\dfrac{\beta_1P_1}{P_Z}\right)+C_{12}\\ R_2\le\Lambda\left(\dfrac{\beta_2P_2}{P_Z}\right)+C_{21}\\ R_1+R_2\le\Lambda\left(\dfrac{\beta_1P_1+\beta_2P_2}{P_Z}\right)+C_{12}+C_{21}\\ R_1+R_2\le\Lambda\left(\dfrac{P_1+P_2+2\sqrt{P_1P_2\bar{\beta}_1\bar{\beta}_2}}{P_Z}\right)\end{array}\right\} \tag{16}$$

**II)** For the case where the additive interference is known causally to both transmitters the following rate region is achievable:

$$\bigcup_{\substack{0\le\beta_1,\beta_2\le1\\ \alpha_i\in\left[-\sqrt{\frac{P_i}{P_S}},\sqrt{\frac{P_i}{P_S}}\right],\\ i=1,2}}\left\{\begin{array}{l}0\le R_1,R_2\\ R_1\le\Lambda\left(\dfrac{\beta_1\tilde{P}_1}{P_Z+(1-\alpha_1-\alpha_2)^2P_S}\right)+C_{12}\\ R_2\le\Lambda\left(\dfrac{\beta_2\tilde{P}_2}{P_Z+(1-\alpha_1-\alpha_2)^2P_S}\right)+C_{21}\\ R_1+R_2\le\Lambda\left(\dfrac{\beta_1\tilde{P}_1+\beta_2\tilde{P}_2}{P_Z+(1-\alpha_1-\alpha_2)^2P_S}\right)\\ \qquad\qquad\qquad\qquad +C_{12}+C_{21}\\ R_1+R_2\le\Lambda\left(\dfrac{\tilde{P}_1+\tilde{P}_2+2\sqrt{\tilde{P}_1\tilde{P}_2\bar{\beta}_1\bar{\beta}_2}}{P_Z+(1-\alpha_1-\alpha_2)^2P_S}\right)\end{array}\right\} \tag{17}$$

where $\tilde{P}_i:=P_i-\alpha_iP_S$.

***Proof of Theorem 4:***

*Part I)* The direct part is derived by substituting the r.v.'s $U,X_1,V_1,X_2,V_2$ defined by (10) in the rate region given by (13). Note that the mutual information functions and the joint p.d.f. of the r.v.'s appeared in the rate region (13) are exactly the same as the rate region (4). Therefore, the consequence of Lemma 1 can be directly applied here. For the converse part it should be noted that the rate region of (16) is the capacity region when there is no interference [5]; hence, it is also an upper bound for the capacity. In fact, the rate region (16) is the capacity when the receiver is also informed of the CSI, i.e., the additive interference $S$.

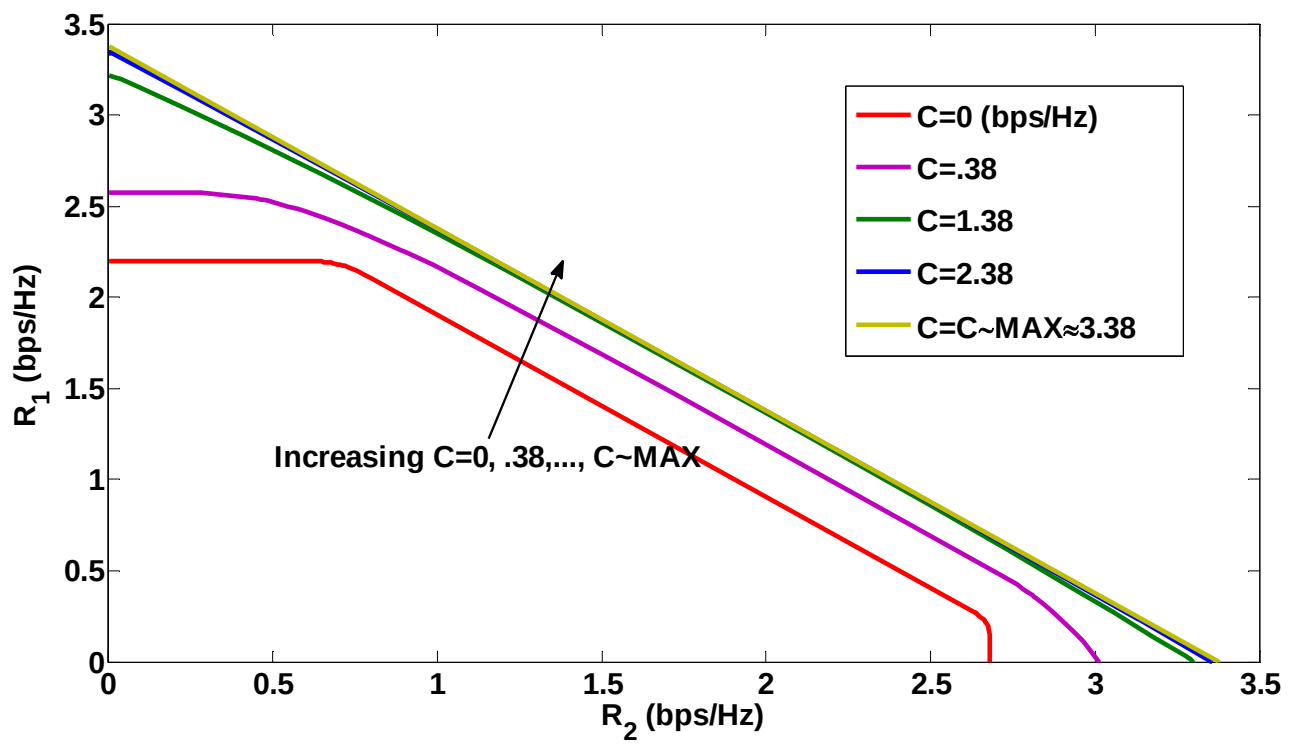

Figure 3. Achievable rate region for the two-user Gaussian MACCE with causal CSI. $P_2 = 2P_1 = 40$, $P_S = 10$, $P_Z = 1$, and $C_{12} = C_{21} = C$.

*Part II)* The achievability of (17) is obtained using the achievable rate region (14) derived for the channel with causal CSI, where the distribution of the r.v.'s is chosen as in Part II of Theorem 2. ■

In Fig. 3, we have plotted the achievable rate region (17) for the two-user Gaussian MACCE with additive interference known casually to both transmitters under different values of $C$, where $C := C_{12} = C_{21}$. As expected, the achievable rate region is enlarged by increasing the value of $C$. However, no further improvement is possible for the values of $C$ greater than $C^* := \max_{0\le\alpha_1,\alpha_2\le 1} \Lambda\left(\frac{\tilde{P}_1+\tilde{P}_2+2\sqrt{\tilde{P}_1\tilde{P}_2}}{P_Z+(1-\alpha_1-\alpha_2)^2 P_S}\right)$, (for the parameters of Fig. 3 this value is about 3.38 (bps/Hz)). Indeed, one can verify that for all values of $C$ equal or greater than $C^*$, the rate region (17) forms a triangle with boundaries determined by the lines $R_1 = 0$, $R_2 = 0$ and:

$$R_1 + R_2 := \max_{0\le\alpha_1,\alpha_2\le 1} \Lambda\left(\frac{\tilde{P}_1+\tilde{P}_2+2\sqrt{\tilde{P}_1\tilde{P}_2}}{P_Z+(1-\alpha_1-\alpha_2)^2 P_S}\right) \tag{18}$$

Therefore, for these values of $C$, no further improvement is possible on the achievable rate region.

# IV. Conclusion

This paper investigates the two-user MAC with cooperative encoders and CSI from an information theoretic point of view, where two different situations are considered: A two-user MACCM and A two-user MACCE. For both situations, the two cases where the CSI is known to the encoders either non-causally or causally are studied. Achievable rate regions are established for the discrete memoryless channels and also the Gaussian channels with additive interference. The achievable rate regions derived for the Gaussian models with additive interference known non-causally to the encoders are shown to coincide with the capacity region of the same channel with no interference, therefore yielding the capacity region for these important cases. The extension of the results to the channels with partial CSI at the encoders would be an interesting next step.

**R. K-Farsani**, (1986, Farsan, I.R. Iran) received the Double Major Bachelor of Science Degree in electrical engineering and pure mathematics from Sharif University of Technology, Tehran, Iran, in July 2010.

He began researching on the *Mathematical Foundations of Information Networks* from June 2007. Up to now, he has been published several papers at the prestigious conferences related to the Information Theory.

He also received the Silver Medal from the Iranian Students Mathematical Olympiad, a country-wide competition, in 2003. From Sep. 2009, he has also been a research scientist with the Radio Communication Group, Education and Research Institute for ICT (ERICT), Tehran, Iran.

**F. Marvasti**, received his BS, Ms and PhD degrees all from Renesselaer Polytechnic Institute in 1970, 1971 and 1973, respectively. He has worked, consulted and taught in various industries and academic institutions since 1972. Among which are Bell Labs, University of California Davis, Illinois Institute of Technology, University of London, King’s College. He was one of the editors and associate editors of IEEE Trans on Communications and Signal Processing from 1990-1997. He has about 60 Journal publications and has written several reference books. His last book is on Non-uniform Sampling: Theory and Practice by Kluwer in 2001. He is also a guest editor on Special Issue on Non-uniform Sampling for the Sampling Theory & Signal and Image Processing journal.  Dr Marvasti is currently a professor at Sharif University of Technology and the director of Advanced Communications Research Institute (ACRI).